\documentclass[showpacs,amssymb,twocolumn,aps,nofootinbib]{revtex4}
\usepackage{amsmath}
\usepackage{amstext}
\usepackage{amsopn}
\usepackage{amsfonts}
\usepackage{amssymb}
\usepackage{bbm}
\usepackage{accents}
\usepackage{empheq}
\usepackage{graphicx}
\usepackage{epsf}
\usepackage{graphics}
\usepackage[latin1]{inputenc}

\begin{document}

\title{Derivation of the fluctuation-dissipation theorem from unitarity}

\author{Ashok K. Das$^{a,b}$ and J. Frenkel$^{c}$}
\affiliation{$^a$ Department of Physics and Astronomy, University of Rochester, Rochester, NY 14627-0171, USA}
\affiliation{$^b$ Saha Institute of Nuclear Physics, 1/AF Bidhannagar, Calcutta 700064, India}
\affiliation{$^{c}$ Instituto de Física, Universidade de São Paulo, 05508-090, São Paulo, SP, Brazil}

\begin{abstract}
Using the closed time path formalism in thermal field theory, we give a derivation of the fluctuation-dissipation theorem which is based on the unitarity of the $S$-matrix.
\end{abstract}

\maketitle
\section{Introduction}
\label{sec:I}

The fluctuation-dissipation (FD) theorem \cite{callen,kubo,zwanzig,simons} is a  powerful tool in the study of  equilibrium phenomena and has proved useful in a variety of circumstances. It relates two seemingly different phenomena in the system, namely, energy dissipation in the medium and the statistical fluctuation of dynamical variables. The effect manifested itself in an unexpected manner in the understanding of the Brownian motion of a particle as well as of the thermal noise in a conductor. In the case of the Brownian motion Einstein had shown that the coefficient of diffusion was related to the coefficient of friction through a temperature dependent factor. Later Nyquist explained theoretically the experimental result of Johnson that, in the absence of an applied current, the mean-square voltage in a conductor is related to the resistance of the conductor. Subsequently the FD theorem has been derived algebraically, starting from the first principles of equilibrium statistical mechanics (as well as in various other ways \cite{chaturvedi, frenkel, millington}). However, the symmetry principle or the conservation law, underlying such a powerful result, has not yet been fully clarified. Einstein had already noted qualitatively that one can understand the relation as follows. When a Brownian particle is subjected to a random force, the same force leads to two components - a statistical fluctuation and a drag - which must, therefore, be related. In this letter, we will show that the FD theorem can be understood as a consequence of the unitarity of the theory. Namely, we will derive the FD theorem starting from the unitarity of the $S$-matrix.

FD theorem holds in equilibrium, both classically as well as quantum mechanically. Here we will describe the quantum mechanical case setting $\hbar = 1$ for simplicity. Let us recapitulate very briefly the algebraic derivation of the FD theorem within the context of a real scalar field theory. The derivation can be generalized to any other field theory in a straightforward manner. There are two essential components in the proof of the FD theorem. The first (and the main) ingredient is that there is a relation between the correlated and the retarded propagators (Green's functions) of the theory. This can be understood as follows. We note that, even though we can define two independent quadratic products from the basic field operators, namely, $\phi (x^{0},\mathbf{x})\phi(0)$ and $\phi(0)\phi(x^{0},\mathbf{x})$ (or, equivalently, the anti-commutator and the commutator), there exists only one independent thermal expectation value or ensemble average (this is also true at zero temperature for the vacuum expectation value). This is easily seen from the relation 
\begin{equation}
\text{Tr} \left(e^{-\beta H} \phi(x^{0},\mathbf{x})\phi(0)\right) = \text{Tr}\left(e^{-\beta H}\phi(0)\phi(x^{0}+i\beta,\mathbf{x})\right).\label{1}
\end{equation}
This is known as the KMS condition \cite{kubo1,schwinger} which follows from the cyclicity of the trace (as well as the identification that the Hamiltonian is the generator of time translation) and we have defined $\beta = \frac{1}{kT}$. Therefore, if we define the correlated and retarded propagators as ($G_{R}(x)$ denotes the retarded Green's function and $\langle\cdots\rangle_{\beta}$ stands for the ensemble average)
\begin{equation}
C (x) = \frac{1}{2}\langle\left[\phi (x), \phi (0)\right]_{+}\rangle_{\beta},\ iG_{R} (x) = \theta(x^{0})\langle\left[\phi(x), \phi(0)\right]\rangle_{\beta},\label{2}
\end{equation}
equation \eqref{1} leads, in momentum space, to the relation for the ensemble averages
\begin{equation}
C(p) = - \coth \frac{\beta p_{0}}{2}\ \text{Im}\, G_{R} (p).\label{3}
\end{equation}
We note here that the negative sign on the right hand side of \eqref{3} is a consequence of the field theoretic definition of $G_{R}(x)$ in \eqref{2}. The second element in the proof of the FD theorem is that, for weak external fields in a linear response theory, one can identify the response function $\chi(p)$ with the retarded Green's function $G_{R}(p)$ of the theory \cite{kubo}. This leads to the result that the statistical fluctuations in a theory in equilibrium given by $C(p)$ are related to the imaginary (dissipative) part of the response function through a temperature dependent factor. This relation is known as the FD theorem.

We note for future use that the time ordered product of two field operators which defines the Feynman propagator of the theory can be written as
\begin{align}
T (\phi (x)\phi(0)) & = \theta (x^{0}) \phi(x)\phi (0) + \theta(-x^{0}) \phi(0)\phi(x)\notag\\
& = \frac{1}{2}\left[\phi(x),\phi(0)\right]_{+} + \frac{\text{sgn}(x^{0})}{2} \left[\phi(x),\phi(0)\right],\label{4}
\end{align}
which, upon taking the ensemble average, leads to
\begin{equation}
iG_{F} (x) = C (x) + \frac{i}{2}\left(G_{R} (x) + G_{R} (-x)\right).\label{5}
\end{equation}
This ties in with the Einstein observation within the field theory context, namely, $C(x)$ and $iG_{R}(x)$ correspond to two parts of the same Feynman propagator and the FD theorem \eqref{3} only gives a relation between these two components. However, it does not explain why this special relation should actually hold. We also note from \eqref{4} that since the anti-commutator is Hermitian while the commutator is anti-Hermitian, it follows that
\begin{equation}
\text{Im}\, G_{F} = - C.\label{6}
\end{equation}
This relation holds in coordinate as well as in momentum spaces since both $\text{Im}\, G_{F}(x)$ and $C(x)$ are real and are even functions of $x$. As a result, we can also write the FD theorem \eqref{3} as
\begin{equation}
\text{Im}\, G_{F}(p) = - C(p) = \coth \frac{\beta p_{0}}{2}\ \text{Im}\, G_{R} (p).\label{7}
\end{equation}
We will use this form of the FD theorem and show that the origin of this powerful relation lies in the unitarity of the theory.

\section{Unitarity and the FD theorem}
\label{sec:II}

In a quantum field theory, conservation of probability is encoded in the unitarity of the $S$-matrix,
\begin{equation}
S^{\dagger} S = S S^{\dagger} = \mathbbm{1},\label{8}
\end{equation}
which implies that the imaginary part of any amplitude in the theory can be given a cutting description. At zero temperature, such a cutting description, even though not required by unitarity, holds graph by graph and is known as the Cutkosky rule \cite{cutkosky}. At finite temperature, it has also been shown  \cite{bedaque,das} to all orders  within the closed time path formalism, that such a description indeed holds for the imaginary part of any  amplitude as a whole (and not graph by graph). 

\begin{figure}[ht!]
\begin{center}
\includegraphics[scale=1]{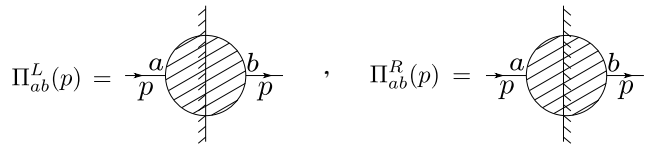}
\end{center}
\caption{Left and right cut self-energy diagrams.}
\label{f1}
\end{figure}

The closed time path formalism \cite{das,landsman}, which we will use, is ideal for the study of nonequilibrium phenomena. It involves doubling the field degrees of freedom (thereby doubling the interaction vertices as well). (We follow the notations and conventions of \cite{das} and make use of many results derived in chapter 5 there in order not to duplicate technicalities.) As a result, the Feynman propagator (Green's function) and the self-energy become $2\times 2$ matrices labelled as $G_{ab}, \Sigma_{ab},\, a,b=\pm$ ($\pm$ denote the doubled thermal degrees of freedom) of the forms (see chapter 2 in \cite{das})
\begin{equation}
G = \begin{pmatrix}
G_{++} & G_{+-}\\
G_{-+} & G_{--}
\end{pmatrix},\quad \Sigma = \begin{pmatrix}
\Sigma_{++} & \Sigma_{+-}\\
\Sigma_{-+} & \Sigma_{--}
\end{pmatrix}.\label{9}
\end{equation}
We note here that the Feynman Green's function $G_{F}$ which we have already defined in \eqref{5} coincides with $G_{++}$. One can also define $2\times 2$ matrices incorporating physical propagators (Green's functions) and self-energies of the forms
\begin{equation}
\widehat{G} = \begin{pmatrix}
0 & G_{A}\\
G_{R} & G_{c}
\end{pmatrix},\quad \widehat{\Sigma} = \begin{pmatrix}
\Sigma_{c} & \Sigma_{R}\\
\Sigma_{A} & 0
\end{pmatrix},\label{10}
\end{equation}
and the two sets of matrices in \eqref{9} and \eqref{10} are related through a $2\times 2$ unitary matrix (see, for example, eqs. (2.42)-(2.43) in \cite{das}). At finite temperature, the results of chapter 5 in \cite{das} show that a cutting description holds, say for example, as a matrix for the self-energy, namely, element by element so that we can write
\begin{equation}
2\,\text{Im}\, \Sigma_{ab} (p) = \Pi_{ab}^{L}(p) + \Pi_{ab}^{R}(p),\quad a,b = \pm,\label{11}
\end{equation}
where $\Pi_{ab}^{L}(p)$ and $\Pi_{ab}^{R}(p)$ represent the cut diagrams for the self-energy with the cut towards the left and right respectively as shown in Fig. \ref{f1}. The retarded self energy is given as the sum 
\begin{equation}
\Sigma_{R}(p) = \Sigma_{++}(p) + \Sigma_{+-}(p).\label{11a}
\end{equation} 
So, for example, for the imaginary parts of the Feynman and the retarded self-energies we can write
\begin{align}
2\,\text{Im}\, \Sigma_{++} (p) & = \Pi_{++}^{L} (p) + \Pi_{++}^{R} (p),\notag\\
2\,\text{Im}\, \Sigma_{R} (p) & = \Pi_{++}^{L} (p) + \Pi_{+-}^{L} (p) + \Pi_{++}^{R}(p) + \Pi_{+-}^{R} (p)\notag\\
& = \Pi_{++}^{L}(p) + \Pi_{+-}^{L}(p),\label{12}
\end{align}
where we have used the result from \cite{bedaque,das} that the right handed cut graphs cancel in the retarded self-energy.
To proceed further with the calculation of the imaginary parts of the self-energies in \eqref{12}, we collect various properties satisfied by the cut diagrams $\Pi_{ab}^{L}(p)$ and $\Pi_{ab}^{R}(p)$. First, we note from Fig. 1 that
\begin{equation}
\Pi_{ab}^{L} (-p) = \Pi_{ba}^{R} (p),\label{13}
\end{equation}
which, in fact, holds graph by graph, but \eqref{13} suffices for our purpose. It is also known that the sum of all the cut diagrams are real, namely,
\begin{equation}
\left(\Pi_{ab}^{L}(p)\right)^{*} = \Pi_{ab}^{L}(p),\quad  \left(\Pi_{ab}^{R}(p)\right)^{*} = \Pi_{ab}^{R}(p).\label{14}
\end{equation} 
Furthermore, there is an underlying symmetry in individual graphs for self-energy at finite temperature, namely, under $(a,b)\leftrightarrow (-a, -b), p\rightarrow -p$ and complex conjugation, any self-energy graph remains invariant (there is a corresponding underlying  symmetry for any graph, not necessarily the self-energy, in a thermal field theory). Together with \eqref{13}, this leads to the result $\Pi_{ab}^{L}(p) = \left(\Pi_{-b,-a}^{R} (p)\right)^{*}$. Equation \eqref{14} then leads to
\begin{equation}
\Pi_{ab}^{L} (p) = \Pi_{-b,-a}^{R} (p),\quad a,b=\pm.\label{15}
\end{equation}
We also use a crucial theorem (proved in chapter 5 of \cite{das}) that the sum over the thermal indices on the cut side of the self-energy diagrams vanishes 
\begin{equation}
\sum_{a=\pm} \Pi_{ab}^{L}(p) = 0 = \sum_{b=\pm} \Pi_{ab}^{R}(p),\label{15a}
\end{equation}
which leads to
\begin{equation}
\Pi_{++}^{R} (p) = - \Pi_{+-}^{R}(p) = - \Pi_{+-}^{L} (p),\label{16}
\end{equation}
where we have used \eqref{15} in the last step. Using the relation \eqref{16}, we can now write\eqref{12} in the simpler form
\begin{align}
& 2\, \text{Im}\, \Sigma_{++}(p) = \Pi_{++}^{L}(p) + \Pi_{++}^{R}(p),\notag\\
& 2\, \text{Im}\, \Sigma_{R} (p) = \Pi_{++}^{L} (p) - \Pi_{++}^{R} (p).\label{17}
\end{align}
 
\begin{figure}[ht!]
\begin{center}
\includegraphics[scale=1]{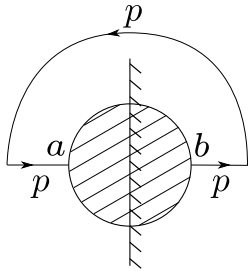}
\end{center}
\caption{Closed loop with a $\Pi_{ab}^{R}(p)$ insertion.}
\label{f2}
\end{figure}

To relate $\text{Im}\,\Sigma_{++}(p)$ and $\text{Im}\,\Sigma_{R}(p)$, we need one final ingredient. Let us consider the closed loop graph shown in Fig. \ref{f2} with a $\Pi_{ab}^{R} (p)$ insertion ($a,b$ fixed)
\begin{equation}
I_{ab} = \int d^{4}p\, \Pi_{ab}^{R} (p) G_{a,-a}(-p),\label{18}
\end{equation}
where we have used the property of thermal graphs (see eq. (5.41) in \cite{das}) that when one end of a propagator is in the cut side of a graph, it is completely determined by the thermal index of the uncut vertex. Furthermore, if we let $p\rightarrow -p$ in the integral and use \eqref{13}, this leads to
\begin{align}
I_{ab} & = \int d^{4}p\, \Pi_{ba}^{L} (p) G_{a, -a} (p)\notag\\
& = \int d^{4}p\,\Pi_{ba}^{L}(p) e^{-a\beta p_{0}} G_{a,-a}(-p).\label{19}
\end{align}
Here we have used the fact that when a thermal propagator has two opposite thermal indices, it satisfies (see eq. (5.43) in \cite{das})
\begin{equation}
G_{a, -a} (p) = e^{-a\beta p_{0}} G_{a, -a}(-p),\quad a=\pm,\label{20}
\end{equation}
which follows from the KMS condition. Comparing \eqref{18} and \eqref{19} we obtain the relation
\begin{equation}
\Pi_{ab}^{R} (p) = e^{-a\beta p_{0}}\, \Pi_{ba}^{L} (p),\quad a,b=\pm,\label{21}
\end{equation}
which, in particular, implies that
\begin{equation}
\Pi_{++}^{R} (p) = e^{-\beta p_{0}} \Pi_{++}^{L}(p).\label{22}
\end{equation}
This equation ensures that, in the zero temperature limit, the self-energy diagram with a right cut vanishes when $p_{0} > 0$, as is known from the study of cutting rules at zero temperature. Substituting equation \eqref{22} into \eqref{17} we obtain a direct relation between the imaginary parts  of the Feynman self-energy and the retarded one,
\begin{equation}
\text{Im}\,\Sigma_{++} (p) = \coth \frac{\beta p_{0}}{2}\ \text{Im}\, \Sigma_{R}(p).\label{23}
\end{equation}
This relation follows from the unitarity of the theory and has the same form as \eqref{7}, but holds for the self-energies.

To make connection with the FD theorem, let us recall that the Green's function and the self-energy in \eqref{9} are related as
\begin{equation}
G^{-1}(p) = \left(G^{(0)}(p)\right)^{-1} - \Sigma(p),\label{24}
\end{equation}
where $G^{(0)}$ denotes the free (tree level) Green's function of the theory. The $2\times 2$ matrix propagator in \eqref{9} can be simply inverted (in the space of thermal indices) and, with some analysis, \eqref{24} leads to the relation
\begin{align}
\text{Im}\, \Sigma_{++} (p) & = - \frac{1}{\det G (p)}\, \text{Im}\, G_{++} (p)\notag\\
&\qquad +\frac{1}{\det G^{(0)} (p)}\, \text{Im}\, G^{(0)}_{++} (p).\label{25}
\end{align}
In a completely parallel manner, one can determine from \eqref{10} the relation
\begin{equation}
\text{Im}\, \Sigma_{R} (p) = - \frac{1}{\det \widehat{G} (p)}\, \text{Im}\, G_{R} (p)+\frac{1}{\det \widehat{G}^{(0)} (p)}\, \text{Im}\, G^{(0)}_{R} (p).\label{26}
\end{equation}
From the fact that $G (p)$ and $\widehat{G} (p)$ (as well as $G^{(0)}(p)$ and $\widehat{G}^{(0)}(p)$) are related by a unitary matrix (so that their  determinants are the same) and using \eqref{23} together with relation (2.51) in \cite{das} we obtain the FD theorem \eqref{7}, namely,
\begin{equation}
\text{Im}\, G_{F}(p) \equiv \text{Im}\, G_{++} (p) = \coth \frac{\beta p_{0}}{2}\ \text{Im}\, G_{R}(p).\label{27}
\end{equation}
This gives a derivation of the FD theorem starting from the unitarity of the theory. 

\section{Conclusion}
\label{sec:III}
Our derivation shows that the physical content of the fluctuation-dissipation theorem may be understood as arising from the unitarity of the $S$-matrix. As we have noted earlier (see \eqref{6} as well as the discussion following \eqref{3}), this result expresses a general relation between the fluctuating properties of a system in thermal equilibrium and the response of the system to a weak external perturbation. Finally, we would like to comment on the classical limit of our result. In our entire analysis, we have set $\hbar = 1$. Restoring the factors of $\hbar$, the temperature dependent factor in \eqref{27} takes the form
\begin{equation}
\coth \frac{\beta p_{0}}{2}\rightarrow \hbar \coth \frac{\hbar\beta p_{0}}{2}.\label{28}
\end{equation}
In the classical limit, $\hbar\rightarrow 0$, this factor reduces to $\frac{2}{\beta p_{0}}$ which leads to the classical fluctuation-dissipation relation in \eqref{3} or \eqref{27}.

\bigskip

\noindent{\bf Acknowledgments}
\bigskip

 A. D. would like to thank the Departamento de F\'{i}sica Matem\'{a}tica in USP for hospitality where this work was done. This work was supported in part by USP and by CNPq (Brazil).


\begin{thebibliography}{10}

\bibitem{callen} H. B. Callen, T. A. Welton, Phys. Rev. {\bf 83},  34 (1951).

\bibitem{kubo} R. Kubo, Rep. Prog. Phys. {\bf 29},  255 (1966).

\bibitem{zwanzig} R. Zwanzig, {\it Nonequilibrium Statistical Mechanics} (Oxford University Press, New York, 2001).

\bibitem{simons} A. Altland, B. Simons, {\it Condensed Matter Field Theory} (Cambridge University Press, Cambridge, 2010).

\bibitem{chaturvedi} S. Chaturvedi, A. Kapoor, V. Srinivasan,  Z. Phys. B {\bf 57},  249 (1984).

\bibitem{frenkel} J. Frenkel, J. C. Taylor, Phys. Rev. E {\bf 85}, 85 (2012).

\bibitem{millington} P. Millington, A. Pilaftsis,  Phys. Rev. D {\bf 88}, 8 (2013).

\bibitem{kubo1} R. Kubo, J. Phys. Soc. Jpn,{\bf 12},  570 (1957).

\bibitem{schwinger} P. C. Martin, J. Schwinger, Phys. Rev. {\bf 115},  1342 (1959).

\bibitem{cutkosky} R. Cutkosky, J. Math. Phys. {\bf 1},  429 (1960).

\bibitem{bedaque} P. F. Bedaque, A. Das, S. Naik, Mod. Phys. Lett. {\bf A12},  2481 (1997).

\bibitem{das} A. Das, {\it Finite Temperature Field Theory} (World Scientific, Singapore, 1997).

\bibitem{landsman} N. P. Landsman, Ch. G. van Weert, Phys. Rep. {\bf 145},  141 (1987).


\end{thebibliography}
\end{document}